\documentclass[aps,prl,preprint,superscriptaddress]{revtex4-1} 
 
 \usepackage{amsmath,bm}
 \usepackage{mathrsfs}
 \usepackage{amsfonts}
 \usepackage{graphicx}
 \usepackage{setspace} 
 \usepackage{graphicx}
 \usepackage{epstopdf}
 \usepackage{dcolumn}
 \usepackage{amsmath}
 \usepackage{epsfig}
 \usepackage{indentfirst}
 \usepackage{psfrag}
 \usepackage{subfigure}
 \usepackage{amssymb}
 \usepackage{color}
 \usepackage{xcolor}
 \usepackage{siunitx}

 \usepackage{graphicx}
 \usepackage{dcolumn}
 \usepackage{bm}
 \usepackage{natbib}
\usepackage{physics}
\usepackage{natbib}
\usepackage[backref=none,bookmarksnumbered=true,bookmarks=true,bookmarksopen=true,colorlinks=true,
citecolor=blue,linkcolor=blue,anchorcolor=green,urlcolor=blue,unicode=false]{hyperref}

\usepackage{ulem}[normalem] 

\makeatletter

\newcommand\colorsout[1]{\bgroup \markoverwith{\textcolor{#1}{\rule[0.5ex]{2pt}{0.4pt}}}\ULon}

\makeatother
\begin{document}

\title{Charge-tunable Cooper-pair diode}

\author{Jon Ortuzar}
  \thanks{The authors contributed equally to this work}
  \affiliation{CIC nanoGUNE-BRTA, 20018 Donostia-San Sebasti\'an, Spain}
  
\author{Stefano Trivini}
  \thanks{The authors contributed equally to this work}
  \affiliation{CIC nanoGUNE-BRTA, 20018 Donostia-San Sebasti\'an, Spain}

\author{Leonard Edens}
  \affiliation{CIC nanoGUNE-BRTA, 20018 Donostia-San Sebasti\'an, Spain}

\author{F. Sebastian Bergeret}
 \affiliation{Centro de F\'isica de Materiales (CFM-MPC) Centro Mixto CSIC-UPV/EHU, E-20018 Donostia-San Sebasti\'an,  Spain}
\affiliation{Donostia International Physics Center (DIPC), 20018 Donostia-San Sebastian, Spain}

\author{Jose Ignacio Pascual}
  \affiliation{CIC nanoGUNE-BRTA, 20018 Donostia-San Sebasti\'an, Spain}
\affiliation{Ikerbasque, Basque Foundation for Science, 48013 Bilbao, Spain}

\begin{abstract}
\textbf{Superconducting diodes, devices that allow Cooper-pair currents to flow more easily in one direction than the other, are set to become key building blocks for dissipationless electronics. Existing realizations, however, rely on magnetic fields, ferromagnets, or complex heterostructures that hinder integration and scalability. Here we demonstrate a diode effect for Cooper-pairs that arises solely from electron–electron interactions in nanoscale superconducting lead islands. When these islands are driven into the Coulomb blockade regime, Cooper-pair transport occurs through resonant charge states. By tuning the island’s electrostatic environment, we controllably break particle–hole symmetry and induce nonreciprocal supercurrents, thereby achieving a gate-switchable superconducting diode without any external magnetic field. Our approach enables robust rectification of superconducting currents and microwave photoresponse, providing a scalable strategy to superconducting logic devices.}
\end{abstract}
\date{\today}
\maketitle

\newpage

\noindent
Nonreciprocity, the breaking of transport symmetry between opposite directions, is a universal consequence of broken fundamental symmetries in  nature~\cite{Onsanger1931,Casimir1945,David2008,wakatsuki2017a}. 
For instance, a conventional semiconductor diode exhibits nonreciprocal current due to the simultaneous breaking of inversion and charge (i.e., particle–hole~\cite{shockley}) symmetries, which naturally occurs in a p–n junction. 
The polarity-dependent resistance of diodes makes them essential for rectification, signal processing, and light detection \cite{Sze2007,scaff1947}. In the realm of quantum technologies, diode-like functionalities for superconducting quantum circuits are highly desired, as they would provide directional control of Cooper-pair currents with minimal dissipation.

Proposals for nonreciprocal superconducting transport rely on the combined breaking of inversion and time-reversal symmetries~\cite{davydova2022c,wakatsuki2017a}. 
The resulting \textit{superconducting diode effect} (SDE) originates from finite-momentum Cooper pairs~\cite{davydova2022c}, giving rise to directional Josephson transport with unequal critical currents ($I_c^+ \neq I_c^-$). 
Most experimental realizations of the SDE exploit the lack of inversion symmetry in non-centrosymmetric superconductors or engineered heterostructures, where the magnetochiral effect induced by external magnetic fields produces directional supercurrents~\cite{Zapata1996,Zapata1998,zhang2020a,ando2020a,baumgartner2022b,pal2022c,daido2022,pal2022a}. 
However, the requirement of external magnetic fields to break time-reversal symmetry hinders practical device integration. 
Alternative approaches using ferromagnetic elements~\cite{narita2022}, multiterminal geometries~\cite{chiles2023a,graziano2020}, or nonreciprocal Landau–Zener dynamics~\cite{wu2022b,kitamura2020} often introduce additional complexity and limited tunability.

Here, we realize the tunable diode effect in a superconducting system in the absence of external magnetic fields or ferromagnetic layers.
Our approach exploits Coulomb interactions in the superconducting transport of Cooper pairs (CP) through nanoscale Pb islands coupled to proximitized graphene. 
Coulomb blockade in these islands opens energy gaps in the zero-bias Josephson resonance, forming two dissipative resonant channels for CP tunneling that can be tuned electrostatically. 
By gating the islands away from charge neutrality, we induce directional Cooper-pair transport in the Josephson regime, with asymmetries fully controlled by the island’s excess charge. Combined voltage- and current-bias measurements reveal that this nonreciprocity arises from the breaking of particle–hole symmetry, in analogy to conventional diodes. 
We exploit this mechanism for two key diode functionalities, gate-tunable rectification of Cooper-pair currents and microwave photodetection, establishing a scalable pathway for superconducting electronics free from external magnetic-field constraints.
\newpage
\begin{figure*}[t]
        \begin{center}
			\includegraphics[width=0.99\textwidth]{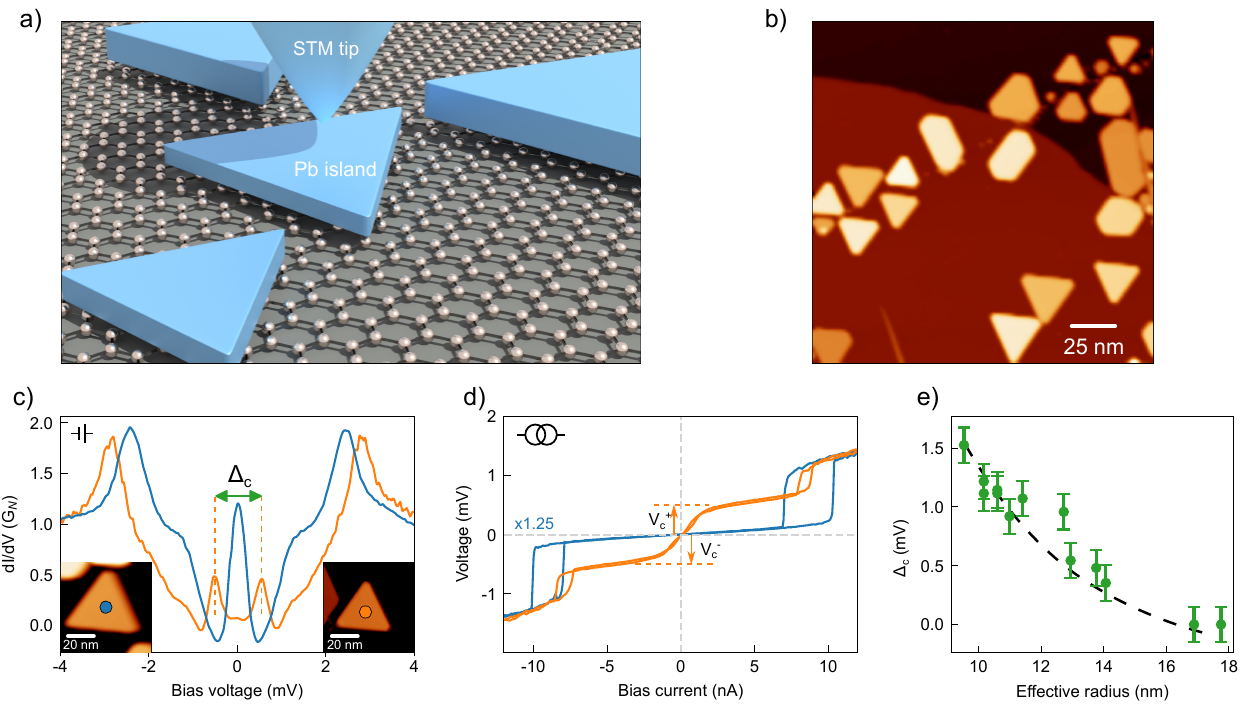}    
        \end{center}
			\caption{ \textbf{Resonant Cooper-pair tunneling in Pb islands on graphene.} (\textbf{a}) Schematic rendering of the experimental setup: an STM superconducting tip approaches a superconducting Pb island on proximitized graphene.
            (\textbf{b}) STM image showing Pb islands of different sizes on graphene ($V$ = 0.4~V, $I$ = 10~pA).            
            (\textbf{c}) Voltage-biased $dI/dV$ spectra of a large island, with effective radious $r_{eff} =\sqrt{Area/\pi}= 18.9$~nm,            
            (blue, $V = 5$~mV, $I = 250$~nA, $R=20$~k$\Omega$), showing a Josephson peak at zero bias,  and a small island, with $r_{eff} = 13.9$~nm   (orange,$V$ = 5~mV, $I$ = 200~nA, $R=25$~k$\Omega$), with symmetric resonant Cooper-pair tunneling (RCT) peaks separated by a voltage gap. Insets: STM images of the two islands.
            (\textbf{d}) Current-biased $V(I)$ characteristics of the islands in panel~\textbf{c}, comparing the zero-voltage Josephson state of the larger (blue) with the voltage step in the smaller (orange), representing the onset for RCT (junction resistances as in \textbf{c}). The orange curve is multiplied by 0.8 to compensate for the different tunneling resistances, allowing comparison of low-resistance plateaus. 
            (\textbf{e}) Dependence of RCT gap on island size, determined from voltage-biased spectra on a set of 12 islands of different areas. The gap increases inversely with effective radius $r_{eff}$, as expected for Coulomb blockade. We identify it as a Cooper pair charging gap, amounting to $4E_C/e$.  }
		\label{fig1}
		\end{figure*}

\textbf{Coulomb blockade in Josephson Currents:}
The superconducting system consists of epitaxial Pb on bilayer graphene (BLG), grown on the silicon-terminated face of a SiC(0001) single crystal (experimental details in supplementary material section S1). At room temperature, characteristic threefold-symmetric islands with typical volumes of $\sim 10^4~\mathrm{nm}^3$ form (\autoref{fig1}b and section S3). The Pb islands are weakly adsorbed on graphene \cite{cortes-delrio2024} and become superconducting at low temperature. 
They induce a uniform superconducting state in the graphene layer, with a gap $\Delta_\mathrm{Gr}$ smaller than that of bulk Pb \cite{Trivini2025a}. Using a low-temperature scanning tunneling microscope (STM) on these islands (\autoref{fig1}a), we created a series of two Josephson junctions (JJs): one between the superconducting Pb-coated STM tip and the island, and another at the island–graphene interface. In our experiments, the tip-island resistance $R_T$ is tunable, enabling control of the junction asymmetry and access to regimes ranging from quasiparticle-dominated tunneling ($R_T \gtrsim 1$ M$\Omega$) to the Josephson regime ($R_T \lesssim  500$ k$\Omega$) \cite{Ruby2015}. The island-BLG junction has a resistance in the order of  1~k$\Omega$, and a capacitance that decreases with the area of their interface.

In \autoref{fig1}c, we compare differential conductance ($\dv*{I}{V}$) spectra taken around 20~k$\Omega$ on a large (blue) and a small (orange) Pb island, corresponding to the colored dots in the insets. The larger island shows the conventional Josephson peak at zero bias, 
originating from the incoherent tunneling of Cooper pairs assisted by the excitation of environmental electromagnetic modes
\cite{averin1990,falci1991,Ingold1992,jack2016,ast2016a,senkpiel2020b}. These processes are described by the probability-of-emission (PoE) function, $P(E)$, which governs the Cooper-pair current $I_S$ \cite{averin1990,falci1991,Ingold1992}:
\begin{equation}\label{poe}
    I_S(V)=\pi e E_J^2\left( P(2eV)-P(-2eV)\right)\; ,
\end{equation}
where $E_J$ is the Josephson coupling energy and $e$ the electron charge (see section S9). In contrast, the smaller island shows no Josephson peak at zero bias. Instead, two symmetric peaks emerge with a gap $\Delta_\mathrm{c}$ between them that increases inversely with their area (\autoref{fig1}e), reaching up to 1.5~mV for islands with an area as small as 100~nm$^2$.

The distinct behavior of small islands is also evident in current-biased measurements, which yield voltage–current traces, $V(I)$, like the ones shown in \autoref{fig1}d (details in section S2). Large islands show the characteristic behavior of phase-trapped JJs, with a nearly zero voltage up to a switching current $\pm I_\mathrm{sw}$, beyond which it jumps into a dissipative quasiparticle branch dominated by Andreev reflections. The traces are hysteretic,  with a retrapping step at $I_r < I_\mathrm{sw}$, as expected for underdamped junctions \cite{barone1982,trahms2023a,steiner2023}. In contrast, small islands display a resistive onset around zero current: the voltage rises with bias until a critical threshold $V_\mathrm{c}$, after which the curve develops a plateau that resembles the low-dissipation Josephson regime but shifted by $V_\mathrm{c}$. Notably, $V_\mathrm{c}$ coincides with the position of the split peaks in \autoref{fig1}d, identifying it as the onset of Cooper-pair tunneling in the presence of Coulomb blockade.




\begin{figure*}[t]
    \begin{center}
	\includegraphics[width=1\textwidth]{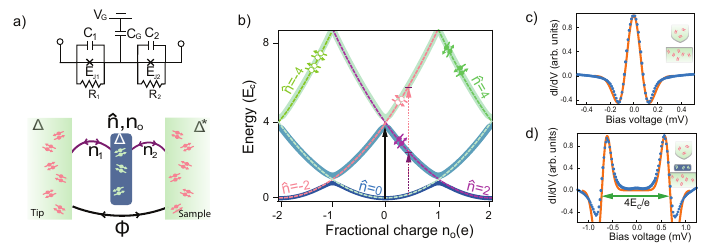}    
    \end{center}
    \caption{\textbf{Josephson effect in the presence of Coulomb blockade.} 
    (\textbf{a}) Circuit model of the double Josephson Junction and, below, scheme of the two mechanisms of Cooper-pair tunneling described in section S10: direct and via (dis)charging the island. 
    (\textbf{b}) Eigen-energies of the charge Hamiltonian~\eqref{charge} as a function of the gate-induced charge offset $n_0$. Parabolic energy bands for each charge state $\hat{n} = -2, 0,$ and $2$ (dashed lines) overlap and open gaps proportional to $E_J$. Vertical arrows mark Cooper-pair tunneling events that add or remove pairs from the island. A finite fractional charge offset $n_0$ breaks particle–hole symmetry for Cooper-pair tunneling (dashed arrows).
    (\textbf{c,d}) PoE simulations (orange) of Josephson dI/dV spectra (blue) of (\textbf{c}) a single and (\textbf{d}) double JJs, using $E_C=0.35$~meV in \textbf{d} and a tip-sample capacitance of 5 fF (1 fF) for the simulation in panel \textbf{c}(\textbf{d}). The PoE in \textbf{d} reproduces RCT peaks spaced by a bias gap amounting to twice the charging energy $4E_C$ divided by 2\textit{e}, the Cooper pair charge, i.e. $4E_c/e$. As the model ignores other transport mechanisms relevant at energies $\sim 4E_C$ (e.g. multiple Andreev reflections, thermal quasiparticle tunneling,..) the negative $\dv*{I}{V}$ dips are overestimated in the simulation.  }
	\label{fig2}
\end{figure*}

\newpage
\textbf{Theoretical modeling:}
In small superconducting islands, Coulomb blockade plays a central role by quantizing the number of Cooper pairs and modifying the excitation spectrum~\cite{Ingold1992}. The system of Fig.~\ref{fig1}a can be modeled as a double Josephson junction (JJ), where the charging, $E_C=\frac{e^2}{C_1+C_2}$ and  Josephson, $E_J$, energies are the key energy scales (Fig.~\ref{fig2}a). The charge term of the Hamiltonian reads 
\begin{equation}\label{charge}
\hat{H}_C = E_C(\hat{n} - n_0)^2 \; ,
\end{equation}
with $\hat{n}$ being the number of paired electrons in the island, $n_0$ is a gate-induced fractional residual charge(see~\autoref{fig2}a and section S10). 
Because $E_C$ grows inversely with capacitance, it becomes appreciable in small islands when $E_C \gtrsim k_B T$, with $T$=1.2~K in our experiment. In this regime, $\hat{n}$ is quantized and the eigen-energies of Eq.~\ref{charge} form parabolic bands as a function of the charge offset  $n_0 = V_G C_G/e$, one for every occupation value $\hat{n}$ (Fig.~\ref{fig2}b) \cite{Ingold1992,bouchiat1998quantum,lafarge_two-electron_1993}.

Josephson coupling hybridizes these charge states, opening gaps at their crossings and allowing the tunneling of Cooper pairs \cite{Ingold1992,joyezphd}. Calculations of the PoE function in the $E_C \gg E_{J1} \gg E_{J2}$ limit, where the charging energies over the Josephson couplings and the $\hat n$ dynamics can be observed, reproduce the suppression of the zero-bias Josephson resonance and its replacement by two spectral resonances at $\pm 2E_C/e$ with the following equation (see Supplementary Material S10 for details):
 \begin{equation}\label{current}
    I=\pi e E_{J1}^2 [P_1(2eV-4E_c(1-n_0))-P_1(-2eV-4E_c(1+n_0)]\; ,
\end{equation}
where, $V$ is the bias voltage and $P_1(E)$ is the PoE distribution seen by the tip-island junction. 
We identify this phenomenon as \emph{resonant Cooper-pair tunneling} (RCT) \cite{joyezphd,vandenBrink1991a,harada1996}.  
In \autoref{fig2}c,d, we compare the dI/dV signals arising from the incoherent tunneling of Cooper pairs in bulk Pb, and a small Pb island, as calculated using Eqs.~\eqref{poe} and~\eqref{current}, respectively~\cite{ast2016a,Ingold_1991}. 

The RCT resonances in \autoref{fig2}d correspond to the environmentally activated tunneling of Cooper pairs, analogous to the Josephson peak between two superconductors, i.e., \autoref{fig2}c. However, unlike standard Josephson tunneling, RCT requires charging transitions (vertical arrows in Fig.~\ref{fig2}b) that supply the energy for pairs to tunnel through the island. Below this threshold, Cooper-pair tunneling is blocked and only the weak coherent transfer of pairs -- $\Phi$ tunneling -- remains (see \autoref{fig2}a and Supplementary Material S10). 

The blockade of Cooper pairs at low voltage also accounts for the resistive response of small islands under low-current bias (Fig.~\ref{fig1}d). 
Once the current induces a voltage drop exceeding the threshold $V_{c} = \pm 2E_C/e$, RCT channels open, giving rise to the low-resistance plateaus observed in the $V(I)$ curves. 
Remarkably, even within this regime of charge quantization, we observe a transition to quasiparticle (Andreev) tunneling with a hysteretic $I(V)$ response, characteristic of underdamped Josephson transport dynamics \cite{barone1982,steiner2023a,trahms2023a}.

\begin{figure*}[t]
    \begin{center}
	\includegraphics[width=0.99\textwidth]{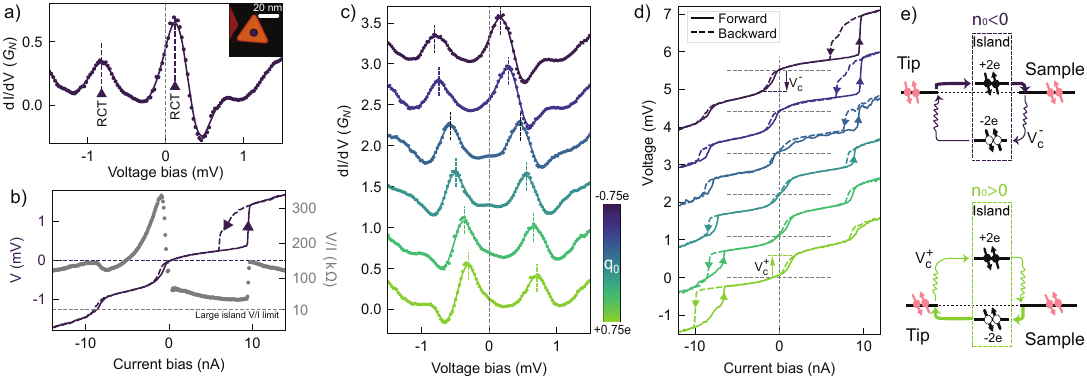}    
    \end{center}
    \caption{\textbf{Gate voltage dependence of the RCT peaks.}
    \textbf{a} dI/dV spectrum of a $r_{eff} = 13.9$~nm Pb island (STM image in the inset) showing voltage-asymmetric RCT peaks. 
    \textbf{b} Current bias V(I) spectrum of the same island in A, asymmetric in resistance and hysteresis. In gray, the calculated V/I resistance shows a pronounced diode effect.
    \textbf{c} Set of dI/dV spectra varying the island excess charge $q_0$ from +0.75e to -0.75e with STM voltage pulses (section S5), inducing a gradual shift of the RCT peaks (V = 5~mV, I = 200~nA, R = 25~k$\Omega$). 
    \textbf{d} Set of V(I) spectrum in the same conditons of \textbf{c}. The diode effect gradually switches polarity with $n_0$ (also shown in linear resistance plots in fig. S6).
    \textbf{e} RCT mechanism in the presence of inversion and particle-hole symmetry breaking. Thicker(thinner) arrows represent the easy(resistive) current polarity. }
	\label{fig3}
\end{figure*}

\textbf{Cooper-pair diode effect:} The Josephson junctions in our device are intrinsically asymmetric ($C_1 \neq C_2$), breaking inversion symmetry. To achieve nonreciprocal transport, an additional symmetry must be lifted. We achieved this by locally breaking particle–hole symmetry through electrostatic gating. As shown in Fig.~\ref{fig2}c, tuning a finite excess charge $n_0$ biases the island toward adding ($n_0>0$) or removing ($n_0<0$) Cooper pairs, turning symmetric tunneling into a directional flow.

In our experiment, a finite amount of $n_0$ is reflected in $dI/dV$ spectra as asymmetric RCT peaks. For example, the spectrum in \autoref{fig3}a shows a stronger resonance slightly above zero bias and a weaker one well below, at negative bias. This imbalance is also well captured by the PoE formalism in the presence of Coulomb correlations, Eq.~\eqref{current}, which assigns distinct environmental channels of Cooper-pair transfer to each bias polarity (fig. S11).

The Cooper-pair diode behavior becomes evident in current-biased measurements like in \autoref{fig3}b: 
Cooper pairs flow readily at positive current bias, while the junction remains resistive for negative currents until a threshold voltage $V_{c}^-\lesssim -4E_C/e$ is exceeded. The easy current polarity exhibits a low voltage plateau and a hysteretic response, both characteristic features of Josephson transport. The resistive branch also shows a plateau above $V_{c}^-$ but with a slightly steeper slope, reflecting increased dissipation.

The fractional charge offset varies from island to island due to electrostatic inhomogeneities from charge puddles at the SiC/graphene interface. Notably, the value of $n_0$ can be tuned on demand over a range of $\pm 1e$ by adjusting the amplitude and polarity of a sequence of voltage pulses applied to the island \cite{Trivini2025}. Analogous to a gate electrode in a conventional three-terminal device, this method enabled us to modulate and even reverse the sign of nonreciprocal Cooper-pair currents.

An example of such gate control is shown in \autoref{fig3}c. Each spectrum was acquired on the same island ($r_{eff}\sim 14$~nm), between a train of voltage pulses applied on top, following the protocol described in section S5~\cite{Trivini2025}. As the excess charge $n_0$ is varied, the RCT peaks shift in energy, effectively moving the Coulomb gap from negative to positive voltage. The displacement of the Cooper-pair resonances around zero bias directly controls the directionality of the Cooper-pair currents. The corresponding $V(I)$ curves in \autoref{fig2}d, directly measured after each $dI/dV$ spectrum, reveal how the diode effect evolves continuously, switching the easy transport direction from positive to negative current bias.  

In agreement with the model of \autoref{fig2}b, the gap between RCT peaks remains fixed at $4E_C/e$ across the explored gate range (Fig.~S7). 
However, their line shape broadens and their amplitude decreases as the corresponding critical bias $V_c$ increases. This behavior is also evident in the current-biased plots of Fig.~\ref{fig3}d: the easy-polarity branches display shallower voltage plateaus, whereas the resistive ones develop steeper slopes (fig. S6). 
The larger dissipation in the hard transport direction likely arises from the mixing of Cooper-pair tunneling with higher-order Zener-like and Andreev processes, increasing with RCT energy. The $V(I)$ curves also show the characteristic hysteretic voltage response near the switching current $I_{sw}\sim10$ nA for the easy direction, which gradually vanishes and disappears as the transport is gated to become dissipative. 
This suggests that Cooper-pair currents in the easy polarity behave similarly to a phase-diffusion transport, which damps in the resistive branch.

\begin{figure*}[t]
    \begin{center}
        \includegraphics[width=\textwidth]{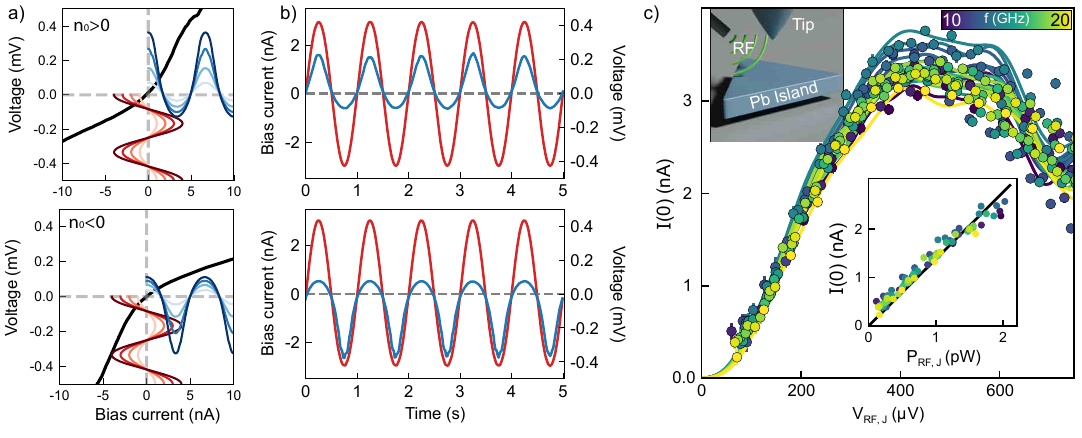}    
    \end{center}
        \caption{\textbf{Cooper pair rectifier behavior.} \textbf{a} $V(I)$ plots of a $r_{eff}=9.5$~nm Pb island in two opposite charge polarities (top, $n_0>0$, bottom $n_0<0$). A simulated sinusoidal current drive around $I=0$ yields an offset non-sinusoidal junction voltage. \textbf{b} Voltage rectification produced by a sinusoidal bias current at drive frequency of \SI{1}{\Hz} applied to each of the two gating configurations of the island in \textbf{a}  (\textit{$V$} = 5~mV, \textit{I} = 250~nA, \textit{R} = 20~k$\Omega$).
        \textbf{c} Detection of microwave radiation using the Cooper pair diode in voltage-biased mode exposed to a radiofrequency antenna. The experimental zero-voltage current $I(0)$ (dots) increases with incident microwave power, independently of frequency (color). Solid lines are fits of Tien-Gordon theory. Error bars correspond to the uncertainty in $I$. Inset shows its dependence on junction power in the small-signal limit, where it is well approximated by a linear fit giving the sensor responsivity. }
		\label{fig4}
\end{figure*}

\textit{Rectification of Cooper-pair currents:} 
The intrinsic non-reciprocity of the gated double JJs system can be tested to rectify the voltage response of Cooper-pair currents, the aimed functionality of a Cooper pair diode. 
Since the rectifying response is strongest when the $V(I)$ characteristics are maximally asymmetric, we prepared a small island ($r_{eff}\sim 10$~nm) with one RCT peak aligned near zero voltage, i.e., close to odd excess charge ($n_0 \rightarrow \pm e$). 
In this regime, the RCT resonance supports the easy flow of Cooper pairs with low dissipation in one direction, while the opposite polarity accounts for most of the voltage drop ($V_C \lesssim 4E_C/e$). When driven by a sinusoidal current, the system generates a rectified (i.e. non-sinusoidal) voltage output, shown in \autoref{fig4}b.  Notably, the rectification polarity can be reversed by gating the islands to opposite residual charge using voltage pulses, as shown in the lower panel of \autoref{fig4}b. 

In our experiments we frequently find voltage rectification ratios, $\text{RR}(I)=V(\pm\SI{1}{\nano\A})/V(\mp\SI{1}{\nano\A})$, amounting to nearly 10 (fig. S8), i.e. within the range of superconducting diodes, spanning from a few to several tens of RR values. Larger values could be possible by reducing the phase-diffusion  resistance of the easy branch at lower temperatures, minimizing environmental dissipation, and increasing the system's capacitance.   


\textit{Microwave photodetector:} The strongly rectifying behavior of the Cooper-pair diode can be exploited for sensing radiation far beyond our detection bandwidth. Figure \autoref{fig4}c plots the current response at zero bias voltage as a function of the microwave amplitude applied to the junction via a nearby antenna, and for different frequencies in the range from \SIrange{10}{20}{\giga\Hz}. A  model based on Tien-Gordon (TG) theory \cite{tiengordon} describes the zero-bias current as multiple photon-assisted CP tunneling (section S11). Naturally, our device is most sensitive when the junction radiofrequency amplitude $V_\mathrm{RF,J}$ is smaller than 2$E_C(1-n_0)$. For larger amplitudes, the $I$-$V$ characteristic becomes increasingly ohmic and current rectification drops (as shown in fig. S9 of supplementary material). Since this zero-voltage current arises from Cooper-pair tunneling, it can be identified as a (photon-assisted) supercurrent. In the small-signal limit, relevant to quantum-metrological applications, we obtain a linear-in-power photodetection responsivity of \SI{1.38}{\nano\A\per\pico\W}, as shown in \autoref{fig4}d. Furthermore, since the  rectification current is given by the junction resistance asymmetry, the responsivity can be finely tuned in-situ via $n_0$, and even inverted (fig. S9).  

In summary, we have demonstrated Cooper-pair diode behavior in nanoscale lead islands, where the nonreciprocal response arises solely from electron–electron interactions in a charge-quantized superconducting island.  Unlike other mechanisms for nonreciprocal superconducting transport, our device operates with smaller current signals (see fig.~S8) and does not rely on an asymmetry of the critical currents ($I_s^+ \neq I_s^-$). The rectification ratio is fully tunable by electrostatic gating, allowing easy integration into superconducting electronics. 
This mechanism, distinct from conventional spin–orbit or magnetochiral routes, establishes charge control as a general pathway for inducing nonreciprocal superconductivity.
Our results open avenues toward scalable, low-dissipation superconducting logic elements and quantum detectors operating without magnetic fields.

\begin{acknowledgments}
We acknowledge financial support from grants 
No. PID2022-140845OBC61, No.  CEX2020-001038-M, No. PID2020-112811GB-I00, and  PID2020-114252GB-I00, funded by MCIN/AEI/ 10.13039/501100011033, 
from the Diputación Foral de Guipuzcoa, 
from the IKUR Strategy of the Department of Education of the Basque Government, and 
from the ERC-AdG CONSPIRA (101097693) funded by the European Union HE program. 
J.O. acknowledges support from the Bask Government through scholarship with No. PRE\_2021\_1\_0350.   
\end{acknowledgments}

\bibliographystyle{apsrev4-1} 
\bibliography{biblio}

\end{document}


\title{Supplementary Material: Charge-tunable Cooper-pair diode }

\author{Jon Ortuzar}
  \thanks{The authors contributed equally to this work}
  \affiliation{CIC nanoGUNE-BRTA, 20018 Donostia-San Sebasti\'an, Spain}
  
\author{Stefano Trivini}
  \thanks{The authors contributed equally to this work}
  \affiliation{CIC nanoGUNE-BRTA, 20018 Donostia-San Sebasti\'an, Spain}

\author{Leonard Edens}
  \affiliation{CIC nanoGUNE-BRTA, 20018 Donostia-San Sebasti\'an, Spain}

\author{F. Sebastian Bergeret}
 \affiliation{Centro de F\'isica de Materiales (CFM-MPC) Centro Mixto CSIC-UPV/EHU, E-20018 Donostia-San Sebasti\'an,  Spain}
\affiliation{Donostia International Physics Center (DIPC), 20018 Donostia-San Sebastian, Spain}

\author{Jose Ignacio Pascual}
  \affiliation{CIC nanoGUNE-BRTA, 20018 Donostia-San Sebasti\'an, Spain}
\affiliation{Ikerbasque, Basque Foundation for Science, 48013 Bilbao, Spain}

\onecolumngrid\

\setstretch{1.1} 
\maketitle

\vspace{2cm}
\tableofcontents
\date{\today}

\setstretch{1.0} 

\onecolumngrid\newpage

\section*{Materials and Methods}

\subsection{Experimental details}
Our experiments have been done in graphene layers grown on Si-terminated SiC(0001) crystals by annealing it to high temperatures \cite{norimatsu2014}. Graphene grows on this side as an electron-doped bilayer with a low density of graphene boundaries. We deposited lead ($\text{T}_c =$ \SI{7.2}{K}) by thermal sublimation on the clean graphene surface at room temperature and under ultra-high vacuum conditions. As shown elsewhere \cite{trivini2025a}, a small amount of lead islands grown on this surface suffices to induce a collective superconducting state, observed as a small gap in the density of states. 

Our experiments are done on a Joule-Thomson low-temperature scanning tunneling microscope from Specs GmbH, at a base temperature of 1.2 K. We acquired differential conductance dI/dV spectra using Pb superconducting tips made by strong indentations of a clean W tip on a clean Pb(111) crystal. All $\dv*{I}{V}$ spectra reported in this paper are measured using a locking amplifier, using ac modulation of the sample voltage of 50 $\mu V$. Current-biased spectra were acquired as described in section S2.     

For the experiments using microwaves, we used an antenna as an emitter, fed with a signal from a microwave generator through a semirigid coaxial cable. The exposed part of the antenna lies at barely 1 cm from the tip-sample junction. The frequency transmission function of the coaxial cables was first determined from $\dv*{I}{V}$ spectra of the SIS tunneling gap (S=superconductor, I=insulator) as a function of radiation frequency, where photon-assisted tunneling results in the split of spectral peaks, following the Tien Gordon model \cite{tiengordon}.  

\subsection{Measurement of current-biased spectra of Josephson Junctions}

\begin{figure}[h]
   \centering
    \includegraphics[width=0.98\linewidth]{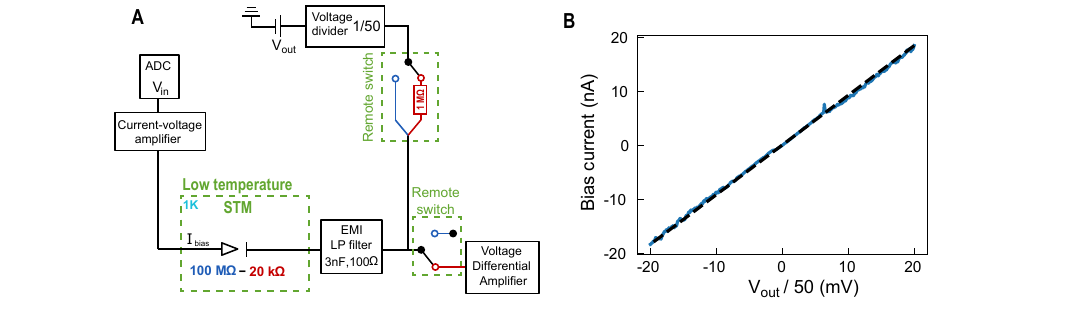}
    \caption{  \textbf{A} Schematics of the circuit for voltage- and current-biased measurements. The remote switch allows us to change from biasing a voltage to biasing a current through the STM junction, while in tunneling regime \cite{trahms2023a}. \textbf{B} Current measured in the current-voltage amplifier in current bias mode as a function of the applied voltage. Using  a 1/50 voltage divider and a a 1~M$\Omega$ resistor, every millivolt of input voltage corresponds to roughly a nanoampere of bias current through the junction. The precise voltage-to-current calibration in every V(I) plot is obtained from these plots. }
    \label{circuit}
\end{figure}

The measurement circuit implemented in this work [\autoref{circuit}A] enables switching between voltage- and current-bias configurations through a 1 M$\Omega$ resistor. To enhance the energy resolution, EMI-type RC filters were included at the bias lines to suppress high-frequency noise. Since the large current-bias resistor couples to the RC filter, thereby reducing the overall frequency cut-off of the circuit, we employed EMI filters with a high cut-off frequency ($>$ 1 MHz). The resulting circuit effectively attenuates high-frequency noise without compromising the measurement bandwidth, thus allowing fast and stable current sweeps.

The voltage drop across the junction is detected with a Stanford Research SR560 voltage preamplifier operated in single-ended mode, with a gain of $10^{3}$ and a 1 kHz low-pass filter. Since the amplifier input can inject high-frequency noise into the junction, it is automatically disconnected by an electronic switch during voltage-bias operation. The current amplifier (FEMTO Messtechnik GmbH DHPCA-100) gives a direct measurement of the real current in the junction (\autoref{circuit}B).

\section*{Supplementary Results}

\subsection{Size and properties of Pb islands on graphene}

\begin{figure}[h]
    \centering
    \includegraphics[width=0.98\linewidth]{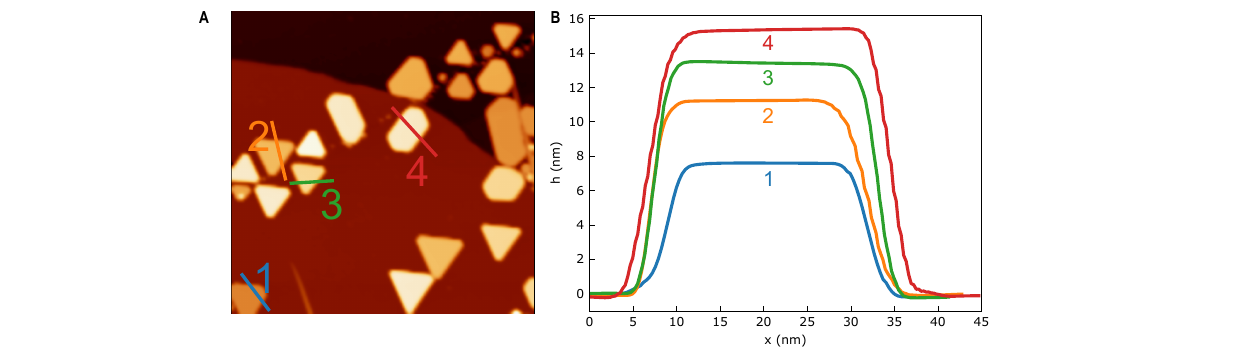}
    \caption{\textbf{A} STM image from Fig.~1 in the main text with line-profiles lines across the edge of four islands. \textbf{B} Plot of the four line profiles for each island. Heights with respect to graphene: 7.60~nm, 11.24~nm, 13.47~nm, 15.34~nm.}
    \label{altura}
\end{figure}

As shown in previous publications \cite{cortes-delrio2021,cortes-delrio2024,Trivini2025,trivini2025a} a (roughly) monolayer amount lead deposited on graphene at room temperature forms flat, polygonal islands  with effective radii $r_{\text{eff}}=(\text{area}/\pi)^{1/2}$ typically ranging from 5~nm to more than 40~nm. The height of Pb islands ranges from 3 nm to 15 nm (\autoref{altura}). The height does not affect the charging energy, which depends on their interface capacitance, proportional to the island area. The islands can be manipulated with the STM tip to different positions over the graphene layer \cite{cortes-delrio2024}.

\clearpage
\subsection {Dependence of Josephson spectra on junction's resistance}

\begin{figure}[h]
    \centering
    \includegraphics[width=0.9998\linewidth]{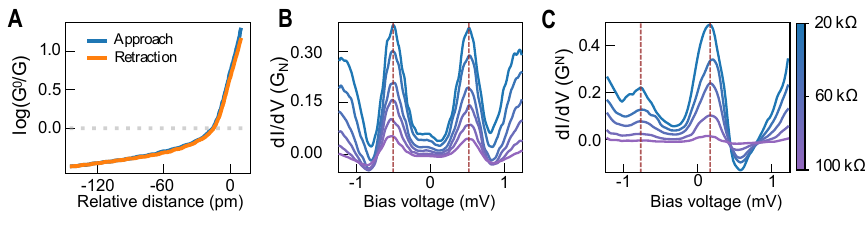}
    \caption{\textbf{Evolution of RCT peaks with junction's resistance.} \textbf{A} junction resistance as a function of tip-island distance. The junction resistance decreases with the relative distance, i.e., with the junction size. \textbf{B} and \textbf{C} resistance-dependent measurements of symmetric and asymmetric RCT peaks, respectively. The vertical dashed lines represent the peak position. }
    \label{fig_sup5}
\end{figure}

Benefiting from the capabilities of the scanning tunneling microscope (STM), we explored the evolution of both the $I(V)$ and $V(I)$ characteristics as a function of the resistance between the tip and the island, $R_1$, by gradually approaching the tip to the island surface. In \autoref{fig_sup5}A, we plot the evolution of the total linear resistance, $R = R_1 + R_2$, with tip approach to a large island (measured with voltage bias values outside the superconducting gap). The resistance decreases exponentially with tip approach up for $z \sim -20$~pm up to point contact at $\frac{1}{G_{o}}$ at around $z \sim -20$~pm. Beyond this point, the slope of the R(z) plot changes. We explore the evolution of RCT peaks and $V(I)$ spectra along this regime, previous to point contact.

\begin{figure}[t]
    \centering
    \includegraphics[width=0.98\linewidth]{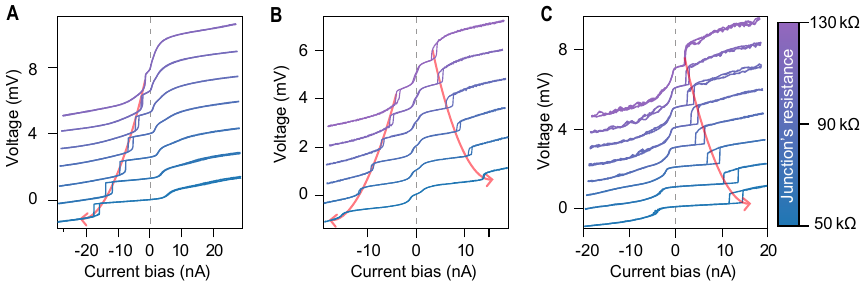}
    \caption{Current bias curves acquired at different tip-sample distances (ranging from 50~k$\Omega$ to 130~k$\Omega$) for three different islands with \textbf{A} $n_0>0$, \textbf{B} $n_0=0$, and \textbf{C} $n_0<0$. The red arrows mark the increase of switching current with the tip's approach towards the island.}
    \label{fig_sup1}
\end{figure}

In \autoref{fig_sup5}\textbf{B} and S3\textbf{C}, we show the evolution of differential conductance spectra with tip approach for a small Pb island. In particular, the sets compare the evolution of RCT peaks for (\textbf{B}) a symmetrically and (\textbf{C}) an asymmetrically gated island, respectively, over a junction-resistance range from 100~k$\Omega$ to 20~k$\Omega$. In both cases, the RCT peaks remain at the same bias positions throughout the tip approach (red dashed lines), while their amplitude increases inversely with the resistance. This behavior is further analyzed in \autoref{DJJ}.

In \autoref{fig_sup1}A–C we compare three tip-approach sets of $I(V)$ characteristics for negatively, neutrally, and positively gated islands, respectively. In these spectra, the switching current, related to the amplitude of the RCT peaks, increases as the tip approaches the islands. For the neutrally charged island (\autoref{fig_sup1}B), the symmetric voltage steps around $I = 0$ do not change height with the approach, in agreement with \autoref{fig_sup5}, and the switching current increases bias symmetrically with junction's conductance (red arrow). For negatively (positively) charged islands, shown in \autoref{fig_sup1}A(C), we find, as expected, hysteretic behavior only in the easy current bias polarity, also with an increasing switching current as the tip approaches.  


\subsection{Gating the electrostatic potential of the Pb island by bias pulsing with an STM} \label{gate}

\begin{figure}[H]
    \centering
        \begin{minipage}[c]{0.6\linewidth}
    \includegraphics[width=1\linewidth]{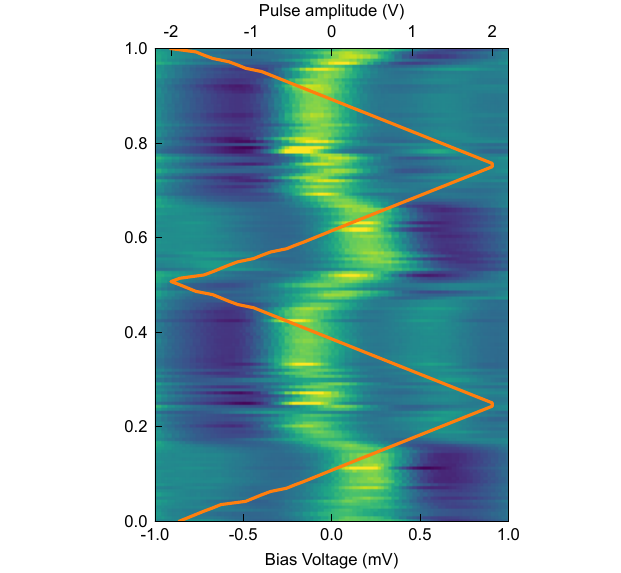}
  \end{minipage}%
  \hfill
  \begin{minipage}[c]{0.35\linewidth}
    \caption{Tunneling spectroscopy on a $r_{\text{eff}} = 12$~nm island as a function of the number and amplitude of applied bias pulses. The orange line represents the amplitude of the bias pulses applied, which is scanned in the range between -2V and 2V.  After every pulse, the shown spectra are measured at the same point on the island. After this procedure, we proved with STM that the island was not modified. 
    }
    \label{gating_RCT}
    \end{minipage}
    \end{figure}

In \autoref{gating_RCT} we show the protocol for manipulating the position of RCT peaks in a $r_{\text{eff}} = 12$~nm Pb island using bias pulses with changing amplitude from –2 to +2~V and back, twice, as indicated by the orange curve. Each pulse was applied with the tip positioned on the island (300~mV, 500~pA, feedback off). For each value of pulse amplitude, a train of five 200~ms pulses was delivered, producing transient currents of –100 to 100~nA. After each pulse sequence, a $\dv*{I}{V}$ spectrum in the Josephson regime was acquired and displayed as a color map in \autoref{gating_RCT}. 
The RCT peaks shift depending mainly on the pulse polarity. Amplitudes above 3.5~V were avoided, as the islands detach and stick to the STM tip.

Importantly, we find that the RCT peaks can be reproducibly tuned to a well-defined level of fractional charging. This controlled state can be reliably obtained in successive experiments, allowing the control of the polarity of the Cooper pair diode. The most significant changes in the spectroscopic features occur when applying pulses in the range $0.8~\text{V} < |V| < 1.2~\text{V}$. While this control range varies somewhat from island to island, in all cases it is possible to adjust and control the spectral asymmetry with high precision.

\subsection{Evolution of RCT peaks and diode resistance with electrostatic gating} 
Figure 3 in the main text shows the evolution of the spectra obtained in a small island as a function of the fractional excess charge. The protocol used to control the excess charge is described in \autoref{gate}. Here, we extract the values of the RCT peaks measured under such control as a function of the excess charge.

\begin{figure}[h]
    \centering
    \includegraphics[width=0.7\linewidth]{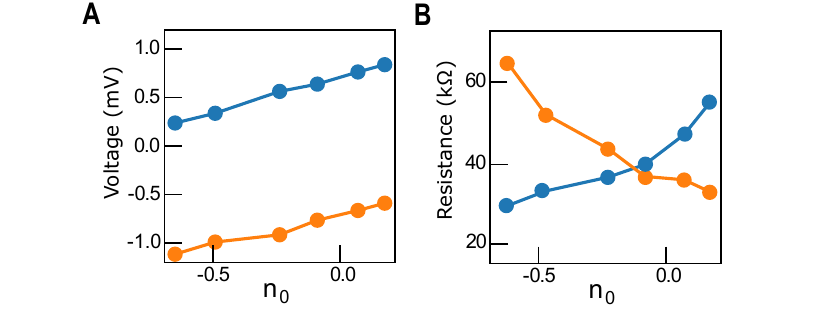}
    \caption{\textbf{A} Position of RCT peaks as a function of excess charge, extracted for the data set shown in Fig 3C as described in the supplementary text. \textbf{B}  Differential resistance of the positive (blue) and negative (orange) voltage plateaus in current-biased $V(I)$ spectra from Fig. 3D, extracted from a linear fit. The slope of the plateaus and, hence, their associated phase-diffusion resistance increases as the peaks move away from zero current.  }
    \label{fig_sup7}
\end{figure}

In \autoref{fig_sup7}A we display the positions of the RCT peaks corresponding to Figure 3C in the main text. These values were obtained from the IV curves rather than from the dI/dV curves, since the maxima of the RCT peaks in the IV curves correspond to the zero-crossing points of the RCT peaks in the dI/dV data. Using the formula $V_{\pm}^{\text{RCT}}=2E_c(n_0\pm1)$ for the positions of the peaks (see \autoref{sec:theory} for details), we have extracted the corresponding excess charge values $n_0$ for each data point. Moreover, we observe that the difference between the positive and negative peaks remains constant, allowing us to determine $4Ec=1.5\pm0.03$~meV. The variance in the peak separation is very small, confirming the robustness of this result.


In Figure 3D of the main text, we show the corresponding current-biased $V(I)$ spectra measured directly after the $dI/dV$ spectra in panel 3C. The critical voltage onset around zero current shift with $n_0$ and accordingly, the constant voltage plateaus above/below at positive/negative bias current. We noted that for easy-polarity (lower critical voltage steps) voltage plateaus were shallower, with their slope apparently increasing with their resistive character. In \autoref{fig_sup7},  we plot the slope of the positive and negative plateaus (i.e. their associated phase-diffusion resistance) as a function of $n_0$, demonstrating that dissipative phase-diffusion resistance increases with the critical voltage onset. As we mentioned in the manuscript, the increasing dissipation of Cooper pair transport with the voltage of its associated RCT channel likely arises from the mixing of Cooper-pair tunneling with higher-order Zener-like and Andreev processes.  
 
In fig. S7 we report the calculated linear resistance plots (V/I) obtained from $V(I)$ plots in Fig. 3D, for their corresponding value of gating. The diode effect becomes evident here, showing a low resistance branch at positive and negative bias current for negative and positive values of $n_0$, respectively. The dissipative branch exhibits a peak of resistance for low currents. The figure shows the gradual change with gating.

\begin{figure}[h]
    \centering  \label{R_gate2}
 \begin{minipage}[c]{0.7\linewidth}
    \includegraphics[width=\linewidth]{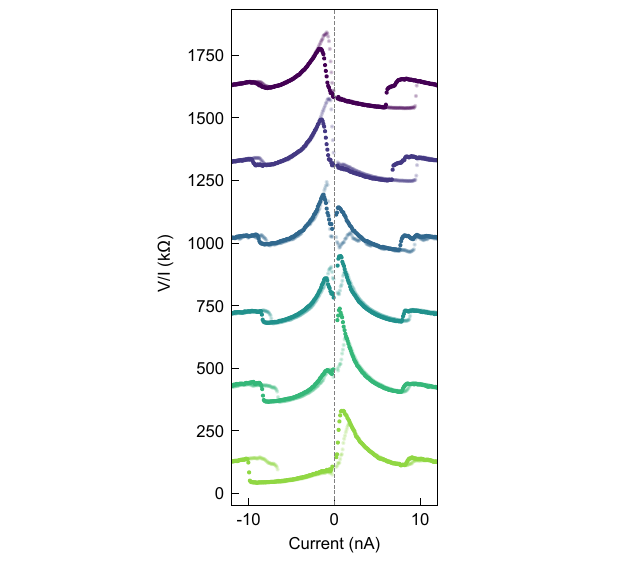}
  \end{minipage}%
  \begin{minipage}[c]{0.3\linewidth}
    \caption{Gate dependence of the linear resistance $V/I$ extracted from the data in Figure 3D of the main text. The forward (solid) and backward (transparent) sweeps highlights the changes in hysteresis around the switching currents. Note how the resistance characteristic inverts by changing the gating. (V = 5~mV, I = 200~nA)
    }
   \end{minipage}
   \end{figure}
\clearpage
\subsection{Voltage and current rectification effect as a function of excess charge}

\begin{figure}[h]
    \centering
    \includegraphics[width=0.78\linewidth]{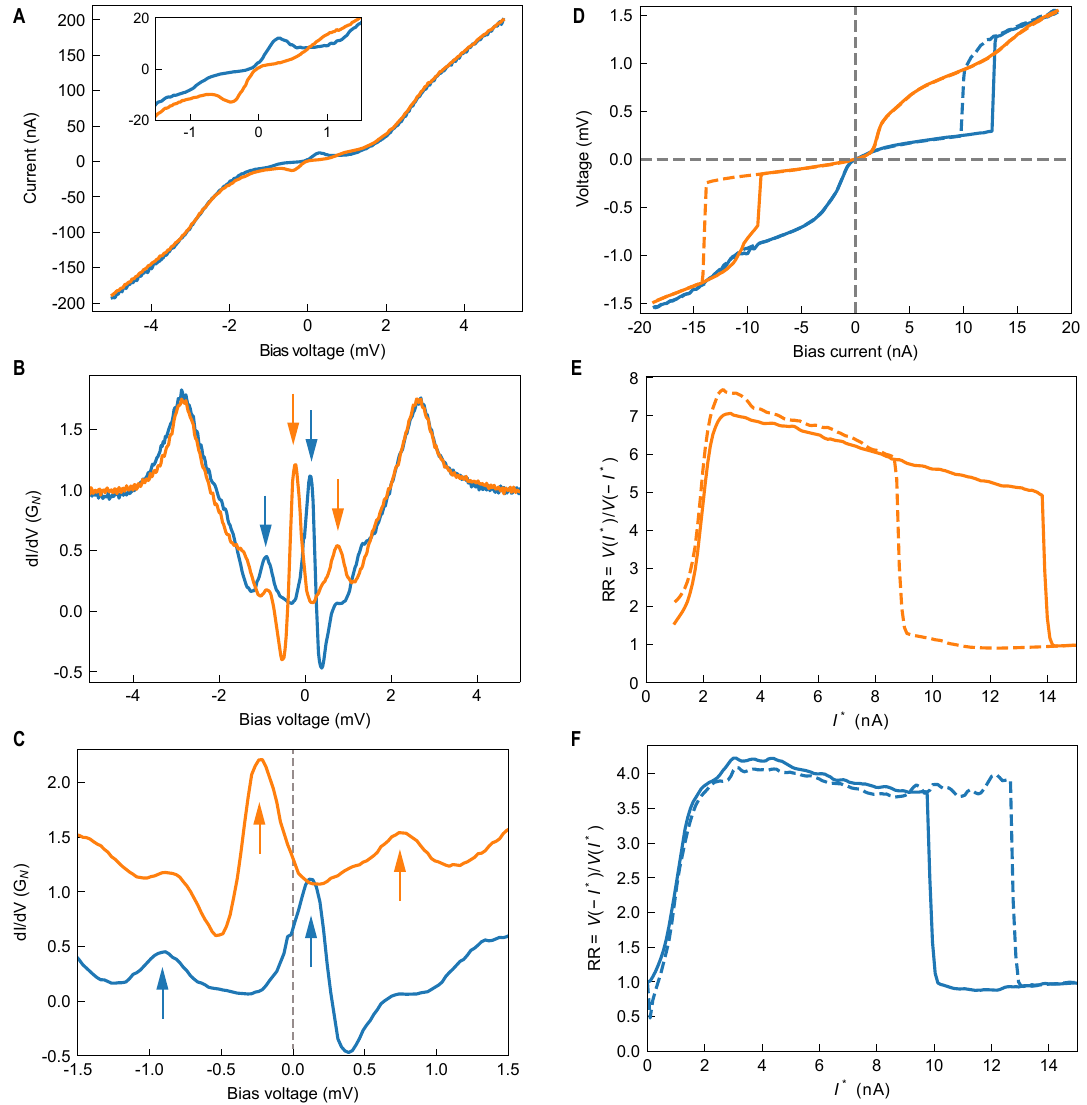}
    \caption{\textbf{Diode behavior and voltage rectification ratio of a a Pb island with two opposite values of residual charge $n_0$.} \textbf{A} Current-voltage characteristics and \textbf{B} its corresponding dI/dV spectrum. \textbf{C} Zoom in the dI/dV spectra of panel \textbf{B} to emphasize the RCT peaks alignment for each $n_0$ and the Coulomb gap between them.   
    \textbf{D} Voltage-current measurements of these two charge configurations. 
    \textbf{E} and \textbf{F} Rectification ratio as a function of the input current.(reference for all spectra V = 5~mV, I = 250~nA)}
    \label{fig_sup8}
\end{figure}

In the main manuscript, the Cooper pair diode effect is tested for two rectification processes, acting either as a voltage rectifier or as a microwave photodetector. In  \autoref{fig_sup8}A-D, we show a set of voltage and current bias spectra of a small island exhibiting two gate states close to $n_0 \lesssim +e$ (orange) and $n_0 \lesssim -e$ (orange), which were controlled by voltage pulses. The asymmetry is maximal. In \autoref{fig_sup8}E-F, we calculated their corresponding voltage rectification ratio $\text{RR}(I)=V(\pm I)/V(\mp I)$ for positive and negative residual charge. As shown there, the RR quickly increases for small currents and remains roughly constant up to the switching current.

 \clearpage

\begin{figure}[h]
    \centering
    \includegraphics[scale=0.65]{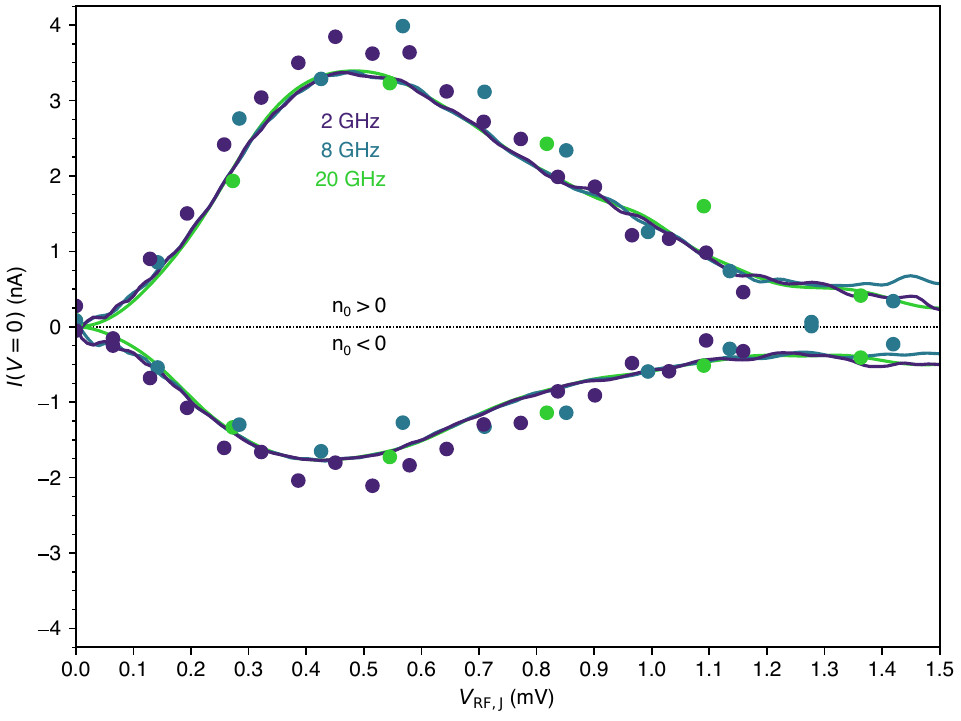}
    \caption{\textbf{RF Rectification effect for positive and negative excess charges at three frequencies.} Dots are measured values, solid lines are fits of \autoref{TG}.}
    \label{fig-rect-q}
\end{figure}

The strongly rectifying behavior of the Cooper-pair diode can be exploited for sensing radiation far beyond our detection bandwidth. Exposing the diode to microwave radiation, the induced voltage oscillation across the junction leads to a rectified current, which we detect as a zero-voltage current. As we show in Fig. 4 of the main text, this current is proportional to the microwave power within a range in the order of the voltages of the Josephson regime in our experiment. In \autoref{fig-rect-q} we show that the reponsivity of the junction can be tuned and reversed by $n_0$ adjusting the stored excess charge. In \autoref{fig-rect-q}, the sign of the zero-bias current flips with $n_0$. This effect arises from the inversion of the nonlinearity at the origin in V-I traces as it rigidly shifts with gating.

\subsection{Estimation of the junction RF power}

To account for the non-sinusoidal current flow through Pb islands, the average junction RF power is obtained from the zero-to-peak voltage amplitude by integration as
\begin{equation}
    P_\mathrm{RF,J}=\frac{1}{T}\int_0^T I(V(t))\cdot V(t)) dt  
\end{equation}
where $V(t)=V_\mathrm{RF,J}\sin{2\pi f t}$ is the incident RF voltage over one period $T$. For simplicity, we neglect reactances introduced by capacitance, such that $P_\mathrm{RF,J}$ is frequency-independent. We note that for large signals, where $I(V)$ becomes linear and $I(t)$ sinusoidal, the power can be simply calculated from the normal state resistance as $P_\mathrm{RF,J}=V_\mathrm{RF,J}^2/2R_\textrm{N}$.

\clearpage
\section*{Theoretical Modeling}\label{sec:theory}

The effects of the environment on the tunneling of Cooper pairs can be described using the Caldeira–Leggett model~\cite{leggett1984}. 
Within this framework, the electromagnetic environment is represented as a collection of harmonic oscillator modes. 
When the interaction with the tunneling electrons is assumed to be weak, its influence can be described starting from the Ambegaokar–Eckern–Schön action~\cite{eckern1984}, 
by modeling the electromagnetic environment through the total effective impedance of the system, $Z$~\cite{leggett1984}. 
In the following, we briefly review the calculation of incoherent tunneling of Cooper pairs in a single junction, expressing the current in terms of the function $P(E)$, 
which gives the probability of exchanging an energy $E$ with the environment.

\begin{figure} [b]
    \centering
    \includegraphics[width=\linewidth]{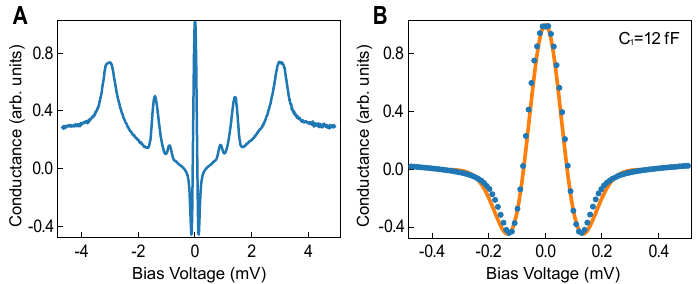}
    \caption{a) Differential conductance measurement between two superconductors: a Pb-coated STM tip and bulk Pb. The coherence peaks appear at $V=\pm 3.5$~mV, the first Andreev reflection at $V=\pm 1.7$~mV, and the second at $V=1.2$~mV. Around zero applied voltage, we measure the Josephson peak. b) Same measurement in a small voltage window around zero voltage showing the Josephson peak (blue curve). The orange curve is the fit obtained following the method in Refs.~\cite{ast2016a} with $C=12$~fF. }
    \label{sup1}
\end{figure}

\subsection{Single Josephson Junction}

The action for a single JJ coupled to an environment can be described by the Ambegaokar–Eckern–Schön action~\cite{eckern1984}

\begin{align}\label{Zphisim}
S = \int \mathcal{D}[\phi]\,
&\exp\Biggl\{
    -\!\int d\tau\, E_J \cos[\phi(\tau)]
    + \!\!\int\!\!\int d\tau\, d\tau'\,
    \alpha(\tau-\tau') \cos\!\left[\frac{\phi(\tau)-\phi(\tau')}{2}\right]
\Biggr\} \nonumber\\
&+ \int d\tau\, \frac{C}{8e^2}(\partial_{\tau}\phi)^2
+ S_z[\phi_z]\; ,
\end{align}

where we disregard the terms with a $\tau+\tau'$ dependence; see Ref.~\cite{joyez2013}.
Here, $C$ is the capacitance of the junction, $\phi_z$ and $\phi$ are the phase differences through the impedance and JJ, respectively, and
\begin{align}\label{albe}
    & \alpha(\tau)=-2|T|^2 g_L(\tau)g_R(-\tau)/2 & & E_J=\int d\tau |T|^2f_L(\tau)f_R(-\tau)\;,
\end{align}
where $g_i(\tau)$ and $f_i(\tau)$ are the imaginary time, real space Green functions projected in the junction. Moreover, the coupling term with the environment reads
\begin{equation}
    S_z[\phi_z]=\dfrac{1}{2\beta}\sum_{\omega}\dfrac{|\omega|}{4e^2}Z(-i|\omega|)\phi_z(\omega)\phi_z(-\omega)\;.
\end{equation}
Here, $Z$ is the impedance of the circuit. The phase across the JJ is related to the voltage drop in the junction through the relation $\partial_{\tau}\phi=2eV$. Note that, due to energy conservation $\phi+\phi_z+\phi_V=0$, where $\phi_V$ is the phase difference in the bias voltage. The supercurrent can be written as
\begin{equation}\label{cur_im}
    \langle I\rangle=E_J \langle \sin{\phi(\tau)}\rangle\; .
\end{equation}

To calculate the phase-phase correlation, we assume that $E_J,\alpha\ll Z^{-1}$. This disregards any crossed interaction between the environment and tunneling current. Assuming a small applied voltage ($V\lesssim\Delta/e$) $\alpha<< E_J,Z^{-1}$, and expanding the partition function to first order in $E_J$, we get the well-known formula for the incoherent tunneling of Cooper pairs~\cite{Ingold1992}
\begin{equation}\label{pprima}
    I=\pi e E_J^2\left(P(2eV)-P(-2eV)  \right)\text{, with } P(\omega)=\int dte^{4J(t)+i\omega t}
\end{equation}
with $J(t)=2\int_0^{\infty} \frac{d\omega}{\omega}\frac{\text{Re}Z_t(\omega)}{R_K}\left\{\coth(\beta\hbar \omega/2)[\cos(\omega t)-1]-i\sin(\omega t)   \right\}$. Here $Z_t(\omega)=(i\omega C+Z^{-1}(\omega))^{-1}$ is the total impedance.

This equation corresponds to Eq. (1) in the main text. 
Figure~\ref{sup1}a shows the dI/dV measurement between a Pb-coated tip and bulk Pb. The peak around zero bias is the Josephson peak, highlighted in Fig.~\ref{sup1}b. The orange curve in Fig.~\ref{sup1}b is obtained by fitting the spectra with the derivative of Eq.~\eqref{pprima}, following the methodology presented in Refs.~\cite{averin1990,Ingold1992,falci1991}.

\subsection{Double Josephson Junction with environment}\label{DJJ}

We now extend the previous single-junction approach to a double-junction system, sketched in Fig.~2a of the main text. The action now includes a new variable associated with the island's degrees of freedom. In this case, the Josephson relations between the time derivatives of the phases and the voltages read
\begin{equation}
    \begin{split}
        & \partial_{\tau}\phi_L(x)=eV_1\\
        & \partial_{\tau}\phi_C(x)=eV_2-eV_1(+eV_G)\\
        & \partial_{\tau}\phi_R(x)=-eV_2,
    \end{split}
\end{equation}
where $V_i$ is the voltage drop in the ith junction, $\phi_j$ is the phase in $j=L,C,R$, and $V_G$ is the gate voltage. We will omit the gate voltage for now. 
It is convenient to change the gauge and write the above expressions as 
\begin{equation}
    \begin{split}
        & \partial_{\tau}(\phi_L(-a)-\phi_C(-a))=\partial_{\tau}\phi_1=2eV_1\\
        & \partial_{\tau}(\phi_C(a)-\phi_R(a))=\partial_{\tau}\phi_2=2eV_2
    \end{split}
\end{equation}
where we assume that $2a$ is the thickness of the island and that the junctions are located at $\pm a$. The phase $\phi_{1,2}$ describes the phase differences across the two junctions. 
The effective action for the double junction system can be obtained from the charging Hamiltonian, given by:
\begin{equation}
    \hat H_C=\dfrac{e^2}{2C_1}\hat n_1^2+\dfrac{e^2}{2C_2}\hat n_2^2+\dfrac{e^2}{2C_{\text{ext}}}(\hat n_1+\hat n_2)\; ,
\end{equation}
where $e \hat n_i$ and $C_i$ are the charge and capacitance of the ith junction, respectively. Moreover, we have added an external capacitance between the left and right leads: $C_{\text{ext}}$. The action for the double junction can be written in the same form as that derived in Ref.~\cite{eckern1984}:

\begin{equation}
\begin{split}
S_0 = \int d\tau\Bigg[ 
    \frac{C_1 V_1^2}{2} + \frac{C_2 V_2^2}{2}\Bigg]
    &= \int d\tau\Bigg[ \frac{C_1}{8e^2}(\partial_{\tau}\phi_1)^2
    + \frac{C_2}{8e^2}(\partial_{\tau}\phi_2)^2
    + \frac{C_{\text{ext}}}{8e^2}(\partial_{\tau}\phi)^2\Bigg] \\
    &=\int d\tau \Bigg[\frac{C_{\Sigma}+C_{\text{ext}}}{8e^2}(\partial_{\tau}\phi)^2
    + \frac{C}{8e^2}(\partial_{\tau}\psi)^2\Bigg]
\; ,
\end{split}
\end{equation}

where two new phases are defined: $\phi=\phi_1+\phi_2$ and $\psi=\kappa_2\phi_1-\kappa_1\phi_2$, with $\kappa_1=C_2/C_{\Sigma}$, $\kappa_2=C_1/C_{\Sigma}$,  $C=C_1+C_2$ and $C_{\Sigma}=C_1C_2/C$. The phase $\phi$ describes the charge transfer through the whole system, while $\psi$ is related to the charge of the island, i.e., $[\hat \phi,\hat k]=i$ and $[\hat \psi,\hat n]=i$, where $\hat n$ and $\hat k$ are the operators used in Eq.~(2) in the main text, so that 
\begin{equation}\label{chargemain}
    \hat H_C=E_C\hat n^2+\tilde E_C \hat k^2\;,
\end{equation}
with $E_C=e^2/C$ and $\tilde E_C=e^2/(C_{\Sigma}+C_{\text{ext}})$. We can now write the total action in the tunneling limit as
\begin{equation}\label{action_double}
\begin{split}
    S=&\sum_{i=1,2}\int d\tau d\tau' \alpha_i(\tau-\tau')\cos{\left(\dfrac{\phi_i(\tau)-\phi_i(\tau')}{2} \right)}+\sum_{i=1,2} E_{Ji}\int d\tau \cos{\phi_i(\tau')}\\
    &+\int d\tau\Bigg[ \dfrac{C_{\Sigma}}{8e^2}(\partial_{\tau}\phi)^2+\dfrac{C}{8e^2}(\partial_{\tau}\psi)^2\Bigg]+S_z[\phi_z]\;,
\end{split}
\end{equation}
where $E_{Ji}$ is the Josephson coupling of the ith junction and $\alpha_i(\tau)$ is defined as in Eq. \eqref{albe} for each junction. As in the previous section, we neglect terms with  $\tau+\tau'$ dependence. The tunneling of Cooper pairs through each junction is described by
\begin{equation}\label{ispoe}
    I_S^i=2eE_{Ji}\langle \sin{\phi_1(\tau)}\rangle\sim 2eE_{Ji}^2\int d\tau \Im{\langle e^{i \kappa_i(\phi(\tau)-\phi(\tau'))}e^{i (\psi(\tau)-\psi(\tau'))}\rangle}\;.
\end{equation}
In principle, one should also include terms related to the direct tunneling of Cooper pairs through the junction, {\it i.e.}, terms proportional to $E_{J1}E_{J2}$. However, we disregard these terms due to the strong asymmetry of the junction and due to its thermal nature. Indeed,  in our setup $E_{J1}\ll E_{J2}$. 
For small voltages and $\alpha\ll E_J\ll Z^{-1}$, both $\phi$ and $\psi$ are separated in the action, so we can calculate their averages independently. The $e^{i\psi}$ is the 2e translator operator for the charge of the island. Then, its average reads
\begin{equation}\label{psi}
    \langle e^{i (\psi(\tau)-\psi(\tau'))}\rangle=\sum_{n,n'}p_n \bra{n}e^{i \psi(\tau)}\ket{n'}\bra{n'} e^{-i\psi(\tau')}\ket{n}=\sum_{n}p_ne^{i(E_n-E_{n-2})(\tau-\tau')}
\end{equation}
Within this approximation, the remaining problem can be solved straightforwardly by following the same steps as in the single-junction case.  The current through the first junction in real time under a dc bias is given by

\begin{equation}\label{psi2}
\begin{split}
I_S^1 = \pi e E_{J1}^2 \int dt \left(
\sum_n p_n e^{2ieV t} e^{i(E_n-E_{n-2})t}
\langle e^{i\kappa_1(\phi(t)-\phi(0))}\rangle \right.\\
\left.
- \sum_n p_n e^{-2ieV t} e^{i(E_n-E_{n+2})t}
\langle e^{i\kappa_1(\phi(t)-\phi(0))}\rangle
\right)\\
= \pi e E_{J1}^2 \sum_n p_n
\left[P_1(2eV+E_n-E_{n-2})-P_1(-2eV+E_n-E_{n+2})\right]
\end{split}
\end{equation}
with $\tilde{P}_i(\omega)=\int dt \exp(4\kappa_i J(t)+i\omega t)$, where $J(t)$ was defined below Eq.~\eqref{pprima}. In the STM setup, having some finite capacitance between tip and ground, as well as sample and ground, is unavoidable. These capacitances effectively enter in $C_{\text{ext}}$, which accounts for the direct capacitance between tip and Graphene as well as the capacitance to ground. Most probably $C_{ext}\gg C_2\gg C_1$, which means that the PoE approximation is valid, i.e., for small applied biases, we are in a mixed regime, where $\hat n$ is well defined, but not $\hat k$. 

For this reason, when the PoE function is calculated for the simulation in Figure~2C,D in the main text, the capacitances we obtain are in the fF regime for both bulk and double junction geometries. The capacitance between tip and substrate does not change much due to the addition of a small island in the. The capacitance between the tip and the "infinite" substrate does not change much due to the tip being a little further away in the case of the double junction, where the islands have a size ranging from 3 to 10~nm.


The probabilities of occupation of a given state of the island, $p_n$ in Eqs.~\eqref{psi} and~\eqref{psi2}, should be calculated by solving a rate equation~\cite{Ingold1992,Hanna1991a}. The tunneling rates in the first and second junctions read
\begin{equation}
\begin{split}
    &\overrightarrow{\Gamma}_{1,n}=\pi eE_{J1}^2P_1(2eV-4E_c(n+1-n_0))\\
    &\overleftarrow{\Gamma}_{1,n}=\pi eE_{J1}^2P_1(-2eV-4E_c(1-n+n_0))\\
    &\overrightarrow{\Gamma}_{2,n}=\pi eE_{J2}^2P_2(2eV-4E_c(n+1-n_0))\\
    &\overleftarrow{\Gamma}_{2,n}=\pi eE_{J2}^2P_2(-2eV-4E_c(n+1+n_0)).
\end{split}
\end{equation}
Then a master equation is used with the defined rates to calculate $p_n$, with even $n$ (see Ref.~\cite{Ingold1992,Hanna1991a}). In the experiment, the Josephson energy of the second junction is fixed and unchanged for any selected island. On the other hand, the Josephson energy of the first junction can be controlled with the tip sample distance. In the measurements presented in the main text, we work in the regime $\Gamma_1\ll\Gamma_2$. Since $\Gamma \propto R^{-2}$ for incoherent tunneling of Cooper pairs, this regime corresponds to $R_1^2 \gg R_2^2$.

In this limit, similar to Ref.~\cite{Ingold1992,Hanna1991a},  the Cooper pair tunneling is fixed by the rates of the first junction. Then, we can write the total current flowing through the junctions as
\begin{equation}\label{current}
    I_n=\pi e E_{J1}^2 (P_1(2eV-4E_c(n+1-n_0))-P_1(-2eV-4E_c(1-n+n_0))
\end{equation}
This expression corresponds to Eq.~(3) in the main manuscript, but it is more general, as it includes, in addition to the fractional residual charge $n_0$, the number of pairs $n$.
 In Fig. 2D in the main text, we plot the differential conductance curve obtained from this expression, for the the particle-hole symmetric case ($n_0$). \autoref{current} captures the main experimental findings, namely two incoherent RCT peaks separated by a voltage by $4E_C/e$ equivalent twice the charging energy of a Cooper pair ($2\times 4E_C$) divided its charge ($2e$). When a finite fractional charge $n_0e$ is added, the peaks shift in energy (fig. S11), reproducing the diode effect discussed in the main text. 
In contrast to experiments, theoretical curves plot positive and negative peaks with the same amplitude because we do not consider effects of charge in the renormalization of the Josephson energy ($E_J$)\cite{matveev1994}, which could effectively be considered by solving the rate equations. 

The model is based on the assumption made that the Josephson coupling of the tip-island junction is smaller than the interface junction. This is well justified considering that point contact measurements on a large island  like in  \autoref{fig_sup5}A, find a contact resistance $R\sim$13~k$\Omega$ for the  double junction in the normal state (voltage bias  50~mV, far from the superconducting gap). Since this resistance $R=R_1+R_2$ is close to the point contact resistance of a single junction $1/G_0\sim 12,9$~k$\Omega$,  the  resistance $R_2$ between the island and Gr should be well below this value. Hence, for the lowest total resistance ($R=R_1+R_2$) used in the experiments, $R\sim$20~kΩ, it is reasonable to stimate that $R_1^2>R_2^2$. 


In \autoref{fig_sup5}B and C we plot the evolution of RCT peaks in symmetrically and asymmetrically gated islands as a function of total resistance. We note that the position of the peaks does not change with tip  approach, but increasing conductance are measured away from the peaks as well as around zero bias. This is probably due to the opening of new tunneling paths through the double junction as the resistance decreases. Still, the simple picture representing charging effects in Eq.~\eqref{current} properly describes the positions of the RCT peaks. 

\begin{figure}[b]
    \centering
\begin{minipage}[c]{0.7\linewidth}
   \includegraphics[width=0.8\linewidth]{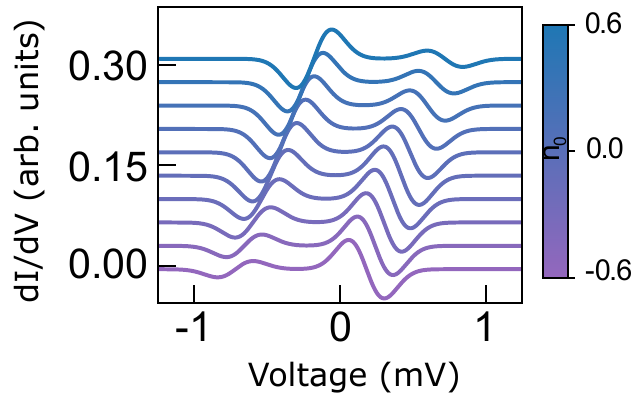}
  \end{minipage}%
  \begin{minipage}[c]{0.3\linewidth}
    \caption{Evolution of the dI/dV of a small island simulated with Eq.~\eqref{current} at different values of $n_0$. We used the same parameters as in Figure 2 in the main text.
    }
   \end{minipage}   
    \label{fig:placeholder}
\end{figure}

\textbf{Simulation of gating:} Using Eq.~\eqref{current} we are able to simulate the evolution with an external gate of a small superconducting island in fig. S11. As expected from Eq.~\eqref{current}, the position of the peaks shifts with $n_0$, which we have taken to span in the same range as the evolution of the island in \autoref{fig_sup1}.

Note that the peak intensities also change with $n_0$. This is something we do not obtain from our effective theory. In order to reproduce this lineshape variations, one has to consider that the $E_{Ji}$ also changes with gating. This has been considered in Refs.~\cite{matveev1994,joyezphd}. In the simulations of \autoref{fig_sup1}, we have included such dependence with gate potential in a phenomenological manner, to demonstrate the similarities with the experiment.  For this, we have considered that $E_{J1}/E_C\sim 0.1$ (temperature has not been included). 

\subsection{Theory of the photodetection}

Following the original paper of Thien and Gordon (TG) \cite{tiengordon}, the effects of photon-assisted tunneling can be accounted for by expanding the total current in terms of N photon-assisted tunneling. The induced current reads~\cite{gonzalez2020}
\begin{equation}\label{eq-tg-curr}
    I(V)=\sum_{n}J^2_n\left(\dfrac{k e V_\mathrm{RF,J}}{\hbar \omega}\right)I^0(V+n\hbar \omega/ke)
\end{equation}
where $V_{RF}$ and $\omega$ are the radiofrequency voltage and frequency, $J_n(x)$ is the $n$th bessel function, $I^0(V)$ is the current without applied radiofrequency and $k$ is the amount of electrones transferred in a elementary tunneling event, i.e., $k=1$ for quasiparticle tunneling and $k=2$ for Cooper pair tuneling. 

Calculating the photodetector sensitivity is not straightforward, and several losses must be considered. The microwave amplitudes referenced in Figure 4C in the main text are obtained from TG theory by relating the splitting of coherence peaks in junctions with large islands as a function of applied junction RF amplitude $V_\mathrm{RF,J}$. This can not be used to calculate the detector's overall sensitivity due to imperfect microwave coupling of the incident radiation $V_\mathrm{RF,0}$ with the tip; we account for these losses by including a multiplicative fit parameter $T(f)$ such that $V_\mathrm{RF,J}=T(f) V_\mathrm{RF,0}$. Yet, the TG formula can be used to calculate the average rate of photon adsorption. By setting $V=0$, the expression for the current reads
\begin{equation}\label{TG}
    I(V=0)=\sum_{n>0}|J_n(V_\mathrm{RF,J}/\Omega)|^2\left[I(n\Omega)+I(-n\Omega)\right]
\end{equation}
with $\Omega=\hbar \omega_{RF}/2e$. Since we are interested in the limit $V_\mathrm{RF}\ll \Delta/e$, where only states inside the superconducting gap have to be considered to calculate the effects of microwave radiation, we set the carrier charge to $2e$. As a consequence, the signal measured in Figure 4C in the main text is a photon-assisted supercurrent flowing in the absence of voltage and magnetic field.


%